\documentstyle[11pt]{article}

\def\openone{%
  \leavevmode\hbox{\xipt1\kern-3.8pt\xiipt1}%
  }

\begin{document}

\begin{titlepage}

\title{Quantum Harmonic Oscillator Algebra and Link Invariants}

\author{C\'esar G\'omez and Germ\'an Sierra \\ Instituto de F\'{\i}sica
Fundamental, Serrano 123, \\Madrid, Spain}

\date{October,1991}

\maketitle

The $q$--deformation $U_q (h_4)$ of the harmonic oscillator
algebra is defined and proved to be a Ribbon Hopf
algebra. Associated with this Hopf algebra we define an infinite
dimensional braid group representation on the Hilbert space of
the harmonic oscillator, and an extended Yang--Baxter system in
the sense of Turaev. The corresponding link invariant is
computed in some particular cases and coincides with the inverse
of the Alexander--Conway polynomial. The $R$ matrix of $U_q
(h_4)$ can be interpreted as defining a baxterization of the
intertwiners for semicyclic representations of $SU(2)_q$ at
$q=e^{2 \pi i/N}$ in the $N \rightarrow \infty$ limit.Finally we
define new multicolored braid group representations and study
their relation to the multivariable Alexander--Conway polynomial.


\end{titlepage}

\section{Introduction}

The connection between quantum groups [1] and link invariants was
first stablished in [2]. The simplest way to describe this
connection is by associating with a given quantum group an
extended Yang--Baxter system in the sense of Turaev [3]. Using
this procedure for the case of $U_q$ (SU(2)) one
reobtains Jone's invariant [4]. Given the finite dimensional
irrep of spin $j$ of $U_q$ (SU(2)), $V^{(j)}$, a realization of
the braid group $B_n$ in End ($\bigotimes^n V^{(j)}$) is defined
in terms of the quantum $R$--matrix of $U_q$ (SU(2)) in the
representation $j$, as follows:

\begin{equation}
\rho: \sigma_i \rightarrow \otimes \cdots \otimes
R^{j,j}_{(i,i+1)} \otimes \cdots \otimes 1
\label{1}
\end{equation}

For any word $\alpha$ in $B_n$ the corresponding link invariant
is defined by:

\begin{equation}
T (\alpha) = a^{- \omega (\alpha)} b^{-n} tr (\rho (\alpha)
\mu^{\otimes n})
\label{2}
\end{equation}

\noindent
where $\mu$ is a diagonal homeomorphism: $\mu: V^{(j)}
\rightarrow V^{(j)}$ satisfying:

\[
(\mu_i \mu_j - \mu_k \mu_{\ell}) R^{k \ell}_{ij} = 0
\]

\begin{equation}
\sum_j R^{kj}_{ij} \mu_j = a b \delta^k_i
\label{3}
\end{equation}

\[
\sum_j (R^{-1})^{kj}_{ij} \mu_j = a^{-1} b \delta^k_i
\]

For the $U_q$ (SU(2)) $-R$ matrix in the representation $j$ the
corresponding extended Yang--Baxter system defined by (3) is
given by

\begin{eqnarray}
\mu & = & q^{H/2}  \nonumber \\
a & = & q^{c_j} (-1)^{2j} \\
b & = & [2j+1] \nonumber
\label{4}
\end{eqnarray}

In the particular case $j = 1/2$ the invariant (2) coincides
with Jone's polynomial. The state model, in the spirit of
Kauffman [5], associated with the invariant (2) can be defined using the
quantum $6j$--symbols of $U_q$ (SU(2)). Moreover the
relationship with vertex integrable models [6] follows by
defining the representation (1) of $B_n$ as the very anisotropic
limit of the trigonometric solution for the Yang--Baxter operators.

In the last few months some new connections relating the
Alexander--Conway polynomial with quantum groups have been
discovered [7--9]. In particular in reference [7] the quantum
group realization of the Alexander polynomial was obtained
starting with the supergroup GL (1,1) which turns out to define
a free fermion model of the invariant. In a completely different
context Date {\it et al} make contact with the Alexander
polynomial in their study of the braid group representations
arising from the chiral Potts model [9]. This result should
indicate a new interplay between the Alexander polynomial and
quantum groups, based on the characterization of the solution to
the start triangle equation for the chiral Potts model in terms
of intertwiners for the cyclic representations of $U_q
(\widehat{SU(2)})$ at roots of unit [10].

In a previous work we have found the intertwiners $R$
matrices for semicyclic representations of $U_q$ (SU(2)) with
$q=e^{2 \pi i/N }$ (for definitions and more details see
references [11, 12]). In the limit of $N$ going to infinity these
$R$ matrices give rise to a new braid group representation
realized on tensor product of an infinite dimensional space
which in fact is isomorphic to the Hilbert space of a harmonic
oscillator [12]. This braid group representation admits a
baxterization [13] which allows the introduction of new parameters
and, interesting enough, the baxterized $R$--matrices satisfies
the Turaev conditions (3) for having a link and knot invariant which
was already defined in [12]. The aim of this paper is twofold:
i)clarify the quantum group origin of this infinite dimensional
braid group representation and ii) give a proper definition and
identification of the link invariant obtained with these new $R$
matrices.

The main results that we obtain are:

\begin{enumerate}
\item The relation of the baxterized $R$--matrices, coming from
the semicyclic irreps of $U_q$ (SU(2)), with the $R$--matrices
of a quantum group deformation of the harmonic oscillator
algebra $h_4$.

\item The identification of the invariant of reference [12] with
the inverse of the Alexander--Conway polynomial.

\item The construction of new colored braid group
representations associated with the quantum group $U_q (h_4)$
and their relation with the multivariable Alexander--Conway polynomial.
\end{enumerate}

These results show that the inverse of the Alexander polynomial
admits a bosonic model in close analogy with the free fermion
model of the Alexander invariant presented in references [7].
The fermionic version of this invariant is based on a quantum
deformation of the supergroup GL (1,1), which in our bosonic
model is replaced by the harmonic oscillator algebra $h_4$.

The plan of the paper is the following: in section 2 we review
the infinite dimensional representations of the braid group
founded in [12]. In section 3 we define the quantum deformation
of the harmonic oscillator algebra $h_4$ and stablish the
connection between the universal $R$ matrix  of this Hopf
algebra and the infinite dimensional braid group representation
introduced in section 2. In section 4 we define the link
invariants and we extend the result to the case of the
multivariable Alexander--Conway polynomial. In Appendix
A we perform some explicit computation of link and knots
invariants, in Appendix B we study the Clebsch--Gordan
decomposition of the irreps of $U_q (h_4)$, and finally in
appendix C we discuss the ribbon structure of $U_q (h_4)$.

\section{The Harmonic Oscillator Braid group representation}

In reference [12] we have considered the $N \rightarrow \infty$
limit $(q= e^{2 \pi i/N})$ of the $U_q$ (SU(2)) Hofp algebra:

\begin{eqnarray}
E F & - & q^2 FE = 1-K^2 \nonumber \\
K E & = & q^{-2} EK \\
KF & = & q^2 FK \nonumber
\label{5}
\end{eqnarray}

\noindent
with the comultiplication:

\begin{eqnarray}
\Delta E & = & E \otimes 1 + K \otimes E \nonumber \\
\Delta F & = & F \otimes 1 + K \otimes F \\
\Delta K & = & K \otimes K \nonumber
\label{6}
\end{eqnarray}

In this limit the Hofp algebra (5) becomes:

\begin{equation}
[E, F] = 1- K^2 \; \; [K,E] = [K,F] = 0
\label{7}
\end{equation}

\noindent
which is isomorphic to the Heisenberg algebra for the harmonic oscillator:

\[
a \equiv \frac{1}{1+k} E \; \; \; \;a^+ \equiv \frac{1}{1-k} F
\]

\begin{equation}
[a, a^+] = 1 \; \; \; \; [k,a] = [k, a^+] = 0
\label{8}
\end{equation}

The semicyclic representations in this limit are now
$\infty$--dimensional and they are labelled by the eigenvalues
of $K$:

\begin{eqnarray}
F e_r & = & (1-\lambda) e_{r+1} \nonumber \\
E e_r & = & r (1+\lambda) e_{r-1} \\
K e_r & = & \lambda e_r \nonumber
\label{9}
\end{eqnarray}

Denoting $H_\lambda$ these irreps, it was proved in reference
[12] that there exist an intertwiner $R^{\lambda_1, \lambda_2} :
H^{\lambda_1} \otimes H^{\lambda_2} \rightarrow H^{\lambda_2}
\otimes {\cal H}^{\lambda_1}$  satisfying the Yang Baxter
equation. The explicit form of the intertwiner is given by

\begin{equation}
R^{r_1 + r_2-\ell,\ell}_{r_1,r_2} (u) = (1+u)^{r_1-\ell} u^{r_2-\ell}
P^{(r_2-\ell, r_1-\ell)}_{\ell} (1-2 u^2)
\label{10}
\end{equation}

\noindent
with $u = \frac{\lambda_1 - \lambda_2}{1 - \lambda_1 \lambda_2}$
and $P^{(\alpha, \beta)}_n (x)$ the Jacobi polynomials:

\begin{equation}
P^{(\alpha,\beta)}_n (x) = \frac{1}{2^n} \; \sum^n_{m=0} \left(
\begin{array}{c} n + \alpha \\ m \end{array} \right) \; \left(
\begin{array}{c} n +\beta \\ n - m \end{array} \right)
(x-1)^{n-m} (x+1)^m
\label{11}
\end{equation}

The Yang--Baxter equation for the $R$--matrix (10) reads:

\begin{equation}
R_i (u) R_{i+1} \left( \frac{u+v}{1+uv} \right) R_i (v) =
R_{i+1} (v) R_i \left( \frac{u+v}{1+uv} \right) R_{i+1} (u)
\label{12}
\end{equation}

The non trivial braid limit of (12) is obtained for $u = v = \pm
1$. In this limit the braid group generators are given by:

\begin{equation}
R^{r'_1 r'_2}_{r_1r_2} (+) = \delta_{r_1+r_2, r'_1+r'_2}
(-1)^{r'_2} 2^{r_1-r'_2} \left( \begin{array}{c} r_1 \\ r'_2
\end{array} \right)
\label{13}
\end{equation}

\begin{equation}
R^{r'_1 r'_2}_{r_1r_2} (-) = \delta_{r_1+r_2, r'_1+r'_2}
(-1)^{r'_1} 2^{r_2-r'_1} \left( \begin{array}{c} r_2 \\ r'_1
\end{array} \right)
\label{14}
\end{equation}

\noindent
which satisfy:

\begin{equation}
R (+) R(-1) = 1
\label{15}
\end{equation}

The representantion of $B_n$ as End $(\otimes^n {\cal H})$ is
defined by:

\begin{equation}
\sigma^\pm_i \longrightarrow 1 \otimes \cdot \cdot R_i (\pm)
\otimes \cdot \cdot 1
\label{16}
\end{equation}

\noindent
with ${\cal H}$ isomorphic to the Hilbert space of the harmonic
oscillator. We will refer to this representation of the braid
group as the harmonic oscillator representation.

A compact way to rewrite eqs (13) and (14) is in terms of the
``universal'' ${\cal R}$ matrix defined as ${\cal R} = PR$ with
$P$ the permutation operator $P (\varphi_1 \otimes \varphi_2 | =
\varphi_1 \otimes \varphi_1 (\varphi_1, \varphi_2 \in {\cal H})$.
Using the creation and annihilation operators $a, a^+$ we find
that $R (\pm)$ admit the following nice representation:

\[
{\cal R} (+) = (e^{i \pi N} \otimes 1) e^{2 a \otimes a^+}
\]

\begin{equation}
{\cal R} (-) = (1 \otimes e^{i \pi N}) e^{2 a^+ \otimes a}
\label{17}
\end{equation}

\noindent
with $N$ the number operator: $N = a^+ a$.

The harmonic oscillator braid group representation (17) admit a
baxterization [13] in the following sense:

\newtheorem{Proposition}{Proposition}[section]
\begin{Proposition}

For arbitrary complex numbers $x,y$ the $R$ matrices:

\[
{\cal R} (x,y,+) = (x^N \otimes y^{-N}) e^{(y-x)a \otimes a^+}
\]

\begin{equation}
{\cal R} (x,y,-) = e^{(x-y)a^+ \otimes a} (y^N \otimes x^{-N})
\label{18}
\end{equation}

\noindent
satisfy the Yang Baxter relation:

\begin{equation}
{\cal R}_{12}(\pm) {\cal R}_{13}(\pm) {\cal R}_{23} (\pm) =
{\cal R}_{23} (\pm)
{\cal R}_{13} (\pm) {\cal R}_{12} (\pm)
\label{19}
\end{equation}

\noindent
where ${\cal R}_{12} (+) = (x^N \otimes y^{-N} \otimes 1) e^{(y-x) a
\otimes a^+ \otimes 1}$, etc.

\end{Proposition}

{}From this proposition if follows that the matrices $R(x,y, \pm)
= P {\cal R}(x,y, \pm)$:

\[
R^{r'_1 r'_2}_{r_1 r_2} (x,y,+) = \delta_{r_1+r_2, r'_1+r'_2}
\left( \begin{array}{c} r_1 \\ r'_2 \end{array} \right)
(y-x)^{r_1-r'_2} x^{r'_2} y^{-r'_1}
\]

\begin{equation}
R^{r'_1 r'_2}_{r_1 r_2} (x,y,-) = \delta_{r_1+r_2, r'_1+r'_2}
\left( \begin{array}{c} r_2 \\ r'_1 \end{array} \right)
(x-y)^{r_2-r'_1} x^{r_2} y^{r_1}
\label{20}
\end{equation}

\noindent
define a $\infty$ dimensional representation of the braid group
$\pi_{x,y}: B_n \rightarrow End  ({\cal H}^{\otimes n})$.
Notice that we recover the $R$--matrices (13), (14) from eq (20)
in the case where $x = -1, \; y=1$.

At a first stage it may appear that we are dealing in (20) with
a two parameter family of inequivalent braid group
representations, but this is not the case since we can prove the
equivalence between the representations $\pi_{x,y}$ and
$\pi_{\alpha x, \alpha y}$ for any $\alpha \neq 0$. This follows
automatically from the relation:

\begin{equation}
R (\alpha x, \alpha y ; \pm) = (\alpha^{-N} \otimes 1) R (x,y)
(\alpha^N \otimes 1)
\label{21}
\end{equation}

Therefore the representation is really characterized by the
ratio $\frac{x}{y}$.

In order to use the braid group representations $\pi_{x,y}$ to
define link invariants we need the following result.

\begin{Proposition}

The braid group representation $\pi_{x,y}$ admits, for arbitrary
$x,y$, an extension a la Turaev.

\end{Proposition}

In fact it is easy to check that for $\mu$ the identity operator
and $a = b^{-1} = \sqrt{(y/x)}$ the set of equations (3) are
satisfied for $R = R(x,y)$ with arbitrary values of $x$ and $y$.
In these conditions the link invariant (2) becomes:

\begin{equation}
T_{x,y} (\alpha) = (x/y)^{1/2 [\omega (\alpha)-n]} tr [\pi_{x,y}
(\alpha)]
\label{22}
\end{equation}

Notice that the invariant only depends on the ratio $x/y$. This
fact follows from the scaling transformations law (21). A proper
way to regularize (22) preserving its invariance under Markov
moves will be defined in section 4.

\section{A quantum deformation of the Harmonic oscillator algebra}

The harmonic oscillator algebra $h_4$ contains four generators
$N, a, a^+, E$ subjected to the relations:

\[
[ N a^+ ]   =  a^+
\]

\begin{equation}
[ N, a ]    =  -a
\label{23}
\end{equation}

\[
[ a, a^+ ]  =    E
\]

\noindent
and with $E$ commuting with all the generators.

This algebra can be obtained as a particular contraction of
U(2). The two casimirs are:

\begin{eqnarray}
c_1 & = & EN - a^+ a \nonumber \\
c_2 & = & E^2 \nonumber
\label{24}
\end{eqnarray}

Next we proceed to define a quantum deformation\footnote{This
quantum deformation should not be confused with the
$q$--oscillators defined in reference [14]}
of $h_4$.

\newtheorem{Definition}{Definition}[section]
\begin{Definition}

The quantum deformed harmonic oscillator algebra $U_q (h_4)$ is
defined by:

\end{Definition}

\[
[a, a^+]  =  \frac{q^E - q^{-E}}{q-q^{-1}}
\]
\[
[N, a]  =  - a
\]
\begin{equation}
[N,a^+]  =  a^+
\label{25}
\end{equation}

\noindent
with the comultiplication:

\begin{eqnarray}
\Delta a & = & a \otimes q^{E/2} + q^{-E/2} \otimes a \nonumber \\
\Delta a^+ & = & a^+ \otimes q^{E/2} + q^{-E/2} \otimes a^+ \\
\Delta E & = & E \otimes 1 + 1  \otimes E \nonumber \\
\Delta N & = & N \otimes 1 + 1 \otimes N \nonumber \\
\label{26}
\end{eqnarray}

The antipode:

\begin{equation}
\gamma (x) = -x \; \; \; x = a, a^+, E,N
\label{27}
\end{equation}

\noindent
and counit:

\begin{equation}
\varepsilon (x) = 0
\label{28}
\end{equation}

\begin{Proposition}

The algebra $U_q (h_4)$ is a quasitriangular Ribbon Hopf algebra.

\end{Proposition}

The universal $R$ matrix satisfying:

\begin{eqnarray}
(\Delta \otimes \; {\rm id}) \; {\cal R} & = & {\cal R}_{13} {\cal R}_{23}
\nonumber \\
({\rm id} \; \otimes \Delta) {\cal  R} & = & {\cal R}_{13} {\cal R}_{12} \\
(\gamma \otimes \; {\rm id}) \; {\cal R} & = & {\cal R}^{-1} \nonumber \\
\Delta' & = & {\cal R} \Delta {\cal R}^{-1} \nonumber
\label{29}
\end{eqnarray}

\noindent
is given by:

\begin{equation}
{\cal R} = q^{-(E \otimes N + N \otimes E)} e^{(q-q^{-1})(q^{E/2} \otimes
q^{-E/2})a \otimes a^+}
\label{30}
\end{equation}

The ribbon structure of $U_q (h_4)$ will be study in section C.

Each irrep of $ U_q
(h_4)$ is labelled by the values of the two Casimirs
$c_1, c_2 (= \frac{q^E-q^{-E}}{q-q^{-1}} N - a^+ a) $
which is equivalent to give the eigenvalues $(e,n)$ of $E$ and
$N$. A generic irrep ($e,n$) is defined as follows:

\begin{eqnarray}
a| r> & = & [e]^{1/2} \sqrt{r} |r-1> \nonumber \\
a^+ |r> & =& [e]^{1/2} \sqrt{r+1} |r+1> \nonumber \\
E |r> & =& e|r> \\
N|r> & = & (r +n) |r> \nonumber
\label{31}
\end{eqnarray}

\noindent
where $\{ |r> \}_{r=0}^\infty$ is a orthonormal basis and the
$q$--number $[e]$ is defined as $[x] = \frac{q^x-q^{-x}}{q-q^{-1}}$.

We shall always consider the cases with $[e] \neq 0$. Evaluating
the universal ${\cal R}$ matrix (30) in the tensor product $(e_1n_1)
\otimes (e_2n_2)$ we define the matrix:

\begin{equation}
R^{e_1e_2} \equiv R(e_1 e_2) = q^{e_1n_2+e_2n_1} P \; {\cal R}^{(e_1
n_1), (e_2n_2)}
\label{32}
\end{equation}

\noindent
where $P$ is the usual permutation operator.In the basis
(31) we get

\[
R^{r'_1 r'_2}_{r_1r_2} (e_1,e_2) = \delta_{r_1 + r_2, r'_1 +
r'_2} \left( \begin{array}{c} r_1 \\ r'_2 \end{array}
\right)^{1/2} \left( \begin{array}{c} r'_1 \\ r_2 \end{array} \right)^{1/2}
\]

\[
\left[ (q^{e_1} - q^{-e_1})(q^{e_2} - q^{-e_2}) \right]^{\frac{r_1-r'_2}{2}}
 q^{\frac{e_1-e_2}{2} (r_1-r'_2)} q^{-e_1 r'_1 -
e_2 r'_2}
\]

\begin{equation}
(R^{-1})^{r'_1 r'_2}_{r_1r_2} (e_1,e_2) = \delta_{r_1 + r_2, r'_1 +
r'_2} \left( \begin{array}{c} r_2 \\ r'_1 \end{array}
\right)^{1/2} \left( \begin{array}{c} r'_2 \\ r_1 \end{array} \right)^{1/2}
\]

\[
\left[ (q^{e_1} - q^{-e_1})(q^{e_2} - q^{-e_2}) \right]^{\frac{r_2-r'_1}{2}}
(-1)^{r_2-r'_1} q^{- \frac{e_1-e_2}{2} (r_1-r'_2)} q^{e_1 r_2 +
e_2 r_1}
\label{33}
\end{equation}

These $R$--matrices satisfy the colored--braid relation:

\begin{equation}
R^{e_2e_3}_{i} \; R^{e_1e_3}_{i+1} \; R^{e_1e_2}_{i}
= R^{e_1e_2}_{i+1}
R^{e_1e_3}_i R^{e_2e_3}_{i+1}
\label{34}
\end{equation}

\noindent
which reduces to the ordinary braid group relation if $e_1 = e_2
= e_3$.

In addition to (34) one has the following properties

\noindent
{\bf P1: Hermiticity}

\begin{equation}
R^{r_1r_2}_{r'_1r'_2} (e_1, e_2) = q^{(e_1-e_2)(r'_1-r_1)}
R^{r'_1 r'_2}_{r_1 r_2} (e_1, e_2)
\label{35}
\end{equation}

\noindent
{\bf P2: Reflection Symmetry:}

\begin{equation}
R^{r'_2r'_1}_{r_2r_1} (e_2 e_1) = (-1)^{r'_1- r_2}
q^{-e_1(r_2+r'_1)-e_2 (r_1+r'_2)}
(R^{-1})^{r'_1 r'_2}_{r_1 r_2} (e_1, e_2)
\label{36}
\end{equation}

Next we would like to compare the $R$ matrices of the quantum
group $U_q (h_4)$ with the matrix $R (x, y, \pm)$ introduced in
the previous section (eqs (20)).

\begin{Proposition}

The $R$ matrices $R(x,y, \pm)$ are equivalent to the $R$ matrix
of $U_q (h_4)$ $R^{\pm 1} (e_1, e_2)$ for $y = x^{-1} = q^{e_1}
= q^{e_2}$.

\end{Proposition}

To prove this proposition we to compare eqs. (20) with
the $R$ matrices given in eqs. (A2, A3), which are related to
eqs (33) by a similarity transformation.

Now from propositions (2.2) and (3.2) it follows that the
matrices $R (e,e)$ and $R^{-1} (e,e)$ satisfy the Turaev
conditions (3) with $\mu$ = id, $a = b^{-1} = q^e$. Consequently the
link invariant (2) defined by the extended Yang Baxter system
associated with $U_q (h_4)$ is the same as the one defined in
section two using the braid group representation $\pi_{x,y}$
with $y/x = q^{2e}$.

\section{The harmonic oscillator link invariants and the Alexander polynomial}

In this section we will proceed to define a regularized version
of (22) which preserves the invariance under Markov moves.

Given an irrep $(e,n)$ of $U_q (h_4)$ the link invariant defined
by the associated extended Yang--Baxter system is:

\begin{equation}
T (\alpha) =  q^{e(-\omega (\alpha) + m)} Tr [\pi_e (\alpha)]
\label{37}
\end{equation}

\noindent
with $\alpha \in B_m$ and $\pi_e (\alpha)$ the infinite
dimensional braid group representation $\pi_{x,y}$ with
$y=x^{-1} = q^e$. By construction $T(\alpha)$ is invariant under
the two Makov moves [3]:

\begin{equation}
\begin{array}{llcl}
(M \cdot I) & T(\alpha \beta) & = & T (\beta \alpha) \\
(M \cdot II) & T (\alpha \sigma_m) & = & T (\alpha) \; \; \alpha
\in B_{m-1} \end{array}
\label{38}
\end{equation}

To make sense of the infinite dimensional trace in (37) we will
proceed to define a regularization similar to the one used in
references [7] [9].

\begin{Definition}

Given $\alpha \in B_m$ such that the corresponding link
$\hat{\alpha}$ is connected, and the irrep $(e,n)$ of $U_q
(h_4)$, we define a regularized trace $T'r (\pi_e (\alpha))$ as follows:

\begin{equation}
T'r (\pi_e (\alpha)) = Tr_{2 \ldots m} (\pi_e (\alpha))
\label{39}
\end{equation}

\end{Definition}

Properly speaking the regularized trace defined by (39) is
associated with the tangle obtained by cutting one strand of the
link $\hat{\alpha}$. If $\hat{\alpha}$ is disconnected we
must, in order to define the regularized trace, to cut one
strand of each component.

\begin{Proposition}

The regularized trace $T'r$ satisfies Markov I.

\end{Proposition}

The proof of this proposition goes as follows. First of all we
must notice that given the tangle obtained by cutting one
strand, let say the $i^{th}$--strand, of the link $\widehat{\alpha
\beta}$ we can always find another strond $i'$ such that the
tangle obtained by cutting the $i'^{th}$ strand of
$\widehat{\beta \alpha}$ is equivalent to the original one. Using
this result, the invariance under Markov I of the regularized
trace (39) reduces to prove that it is independent of which
strand we chose to cut, namely:

\begin{equation}
Tr_{2,\cdot\cdot m} (\pi_e (\alpha)) = Tr_{1, \cdot\cdot i,
i+2..m} (\pi_e (\alpha))
\label{40}
\end{equation}

The identity (40) is a formal consequence of the following lemma.

\noindent
\underline{Lemma 4.1}

\begin{equation}
T'r (\pi_e (\alpha)) \propto \; \openone
\label{41}
\end{equation}

Proof: Using the quasitriangularity of $U_q (h_4) : \Delta' =
{\cal R} \Delta {\cal R}^{-1}$, we obtain:

\begin{equation}
\Delta^{(m)} (a^+) \pi_e (\alpha) = \pi_e (\alpha) \Delta^{(m)} (a^+)
\label{42}
\end{equation}

\noindent
for any $\alpha \in B_m$. Exponentiating equation (42) we get:

\begin{equation}
\pi_e (\alpha) (e^{za^+} \otimes \openone \otimes..) = (e^{za^+} \otimes
\openone \otimes..)(\openone \otimes \Omega) \pi_e (\alpha)
 (\openone \otimes \Omega^{-1})
\label{43}
\end{equation}

\noindent
where $\Omega$ is a similarity transformation. Now defining the
trace in the basis of coherent states: $|z> = e^{za^+}|0>$ and
using (43) we obtain the desired result (41).

Invariance under Markov II is obtained by including the same
prefactor as in (37). At this point we can define a normalized
link invariant by:

\begin{equation}
Z_{\hat{\alpha}} (q^e) =  q^{e(- \omega (\alpha) + m-1)} T'r
(\pi_e (\alpha))
\label{44}
\end{equation}

\noindent
where the normalization is given by:

\begin{equation}
Z(unknot) = 1
\label{45}
\end{equation}

The invariant $Z_{\hat{\alpha}}$ is a polynomial in $q^e$. In
all the examples we have considered (see Apprendix A for some
non trivial cases) we have found the result:

\begin{equation}
Z_{\hat{\alpha}} (q^e) = \frac{1}{\Delta_{\hat{\alpha}}(t)}
\label{46}
\end{equation}

\noindent
with $\Delta_{\hat{\alpha}}(t  = q^e)$ the Alexander
polynomial. The result (46) is very natural in our approach
where we have considered as the starting point the Hopf algebra
$U_q (h_4)$ which can be thought as a bosonic version of GL
(1,1). We do not yet have a general proof of (46).
Notice that although $\Delta_{\hat{\alpha}}$ satisfies a skein
rule its inverse does not. This is of course related to the fact
that the braid group representation underlying
$Z_{\hat{\alpha}}$ is infinite dimensional and that $R(e,e)$
has an infinite number of different eigenvalues for generic values of
$q^e$. This means that the proof of (46) cannot probably proceed
through the skein rule as is usually done for other invariants
coming from quantum groups.

The colored version of the previous invariant is easily obtained
once we have a colored braid group representation as in (34).
Proceeding as the authors of ref [9] we first renormalized the
matrix $R(e_1,e_2)$ according to:

\begin{equation}
\tilde{R} (e_1,e_2) \equiv q^{- \frac{e_1+e_2}{2}} R (e_1, e_2)
\label{47}
\end{equation}

\noindent
such that $\tilde{R} (e_1, e_2)$ do satisfies the braid group
relation while the Turaev conditions (3) holds now with

\begin{equation}
\sum_j \tilde{R} (e,e)^{kj}_{ij} = \sum_j \tilde{R}^{-1}
(e,e)^{kj}_{ij} = q^{-e/2} \delta^k_i
\label{48}
\end{equation}

\noindent
with this redefinition the colored version of the invariant (37)
reads:

\begin{equation}
T (\hat{\alpha}) = q^{\sum^n_{i=1} e_i}  Tr \tilde{\pi}
(\alpha) = \prod^n_{i=1} t_i \; \; Tr \tilde{\pi} (\alpha)
\label{49}
\end{equation}

\noindent
where $\tilde{\pi} (\alpha)$ is the representation of $\alpha$
in terms of the matrix $\tilde{R}$. Notice that (49) reduces to
(37) if $e_1 = e_2 = .. = e_n = e$.

Finally if we normalized the invariant (49) to be equal to
$(t_1-t^{-1}_1)$ for the unknot with color 1,
then it becomes:

\begin{equation}
Z_{\hat{\alpha}} = (t_1 - t^{-1}_1) \prod^n_{i=2} t_i \; \; T'r
\tilde{\pi} (\alpha)
\label{50}
\end{equation}

\noindent
as an exercise one can check that

\begin{equation}
Z_{\hat{\sigma^2_1}} = (t_1 - t^{-1}_1) t_2 \; \;T'r
\tilde{R}^{e_2 e_1} \tilde{R}^{e_1e_2} = 1
\label{51}
\end{equation}

\noindent
which is the correct value for the two component link in the
normalization adopted above.

Here again we conjecture that the invariant (50) is the inverse
of the multivariable Alexander Conway polynomial.

\newpage
\section*{Appendix A}

\subsection*{Calculation of some links and knots invariants}

In this appendix we shall compute some invariants of knots and
links which shall support the conjecture (46), i.e. that the
invariant $Z_{\hat{\alpha}}$ is the inverse of the
Alexander--Conway polynomial.

Instead of using the normalized basis $|r>$ defined in eqs. (31)
it will be more convenient to use the following basis for the
irrep $(e,n)$ of $U_q (h_4)$:

$$
\begin{array}{lcl}
a \;v_r & = & [e] \;r \; v_{r-1} \\
a^+ \; v_r & = &    v_{r+1} \\
N \; v_r & = & (r + n) \; v_r \\
E \; v_r & = & e \;v_r \end{array}
\eqno{(A1)}
$$

The braiding matrices (33) now read:

\[
R^{r'_1 r'_2}_{r_1 r_2} (e_1, e_2) = \delta_{r_1 + r_2, r'_1 + r'_2}
\]

$$
\left( \begin{array}{l} r_1 \\ r'_2 \end{array} \right)
(t_1-t^{-1}_1)^{r_1-r'_2} (t_1 t^{-1}_2)^{\frac{r_1-r'_2}{2}}
t^{-r'_1}_1 \; t^{-r'_2}_2
\eqno{(A2)}
$$

\[
(R^{-1})^{ r'_1 r'_2}_{r_1 r_2} (e_1, e_2) = \delta_{r_1 + r_2, r'_1 + r'_2}
\]

$$
\left( \begin{array}{l} r_2 \\ r'_1 \end{array} \right)
(t^{-1}_2-t_2)^{r_2-r'_1} (t_2 t^{-1}_1)^{ \frac{1}{2}(r_2 - r'_1)}
t^{r_2}_1 \; t^{r_1}_2
\eqno{(A3)}
$$

\noindent
where $t_1 = q^{e_1}$ and $t_2 = q^{e_2}$.

It is clear from (41) that the invariant $Z_{\hat{\alpha}}$
given in eq. (44) can also be compute as:

$$
Z_{\hat{\alpha}} = (1 - \mu) \;  q^{e(m-\omega(\alpha)-1)} Tr
\pi_e (\alpha) (\mu^{N-n} \otimes \openone - \otimes \openone)
\eqno{(A4)}
$$

\noindent
since the trace of $\mu^{N-n}$ in the first space is equal to
$(1-\mu)^{-1}$ for any $\mu \in C (|\mu| < 1)$.

Let us consider the case of the link obtained as
$\widehat{\sigma^2_1}$ with $\sigma_1 \in B_2$. Here we have to
evaluate

\[
Tr (R^{ee} R^{ee} (\mu^{N-n} \otimes \openone)) =
\sum^\infty_{\begin{array}{l} r_1 r_2=0 \\ \ell_1 \ell_2
\end{array}} \; \; R^{\ell_1 \ell_2}_{r_1 r_2} (e,e)
R^{r_1r_2}_{\ell_1 \ell_2} (e,e) \mu^{\ell_1}
\]

$$
= \sum_{r_1 r_2 \ell_1 \ell_2} \; \delta_{r_1 + r_2, \ell_1 +
\ell_2} \left( \begin{array}{l} r_1 \\ \ell_2 \end{array}
\right) \; \left( \begin{array}{l} \ell_1 \\ r_2 \end{array}
\right) \; \; (t - t^{-1})^{r_1-\ell_2 + \ell_1 - r_2}
\eqno{(A5)}
$$

\[
\times t^{-(r_1+r_2 + \ell_1 + \ell_2)} \; \; \mu^{\ell_1}
\]

First we sum $r_1$ in order to eliminate the delta ``function'' obtaining:

$$
\sum_{r_2 \ell_1 \ell_2} \left( \begin{array}{c} -1-\ell_1+r_2
\\ \ell_2 \end{array} \right) \; \; \left( \begin{array}{c}
\ell_1 \\ r_2 \end{array} \right) (-1)^{\ell_2} (t - t^{-1})^{2
(\ell_1 - r_2)} t^{-2 (\ell_1 + \ell_2)} \mu^{\ell_1}
\eqno{(A6)}
$$

\noindent
where we have used the formula:

$$
\left( \begin{array}{c} r\\ n \end{array} \right) = (-1)^n
\left( \begin{array}{c} -1 + n -r \\ n \end{array} \right)
\eqno{(A7)}
$$

\noindent
to rewrite the combinatorial number $\left( \begin{array}{c} r_1
\\ \ell_2 \end{array} \right)$. Now (A6) can be computed
performing first the sum over $\ell_2$, next over $r_2$ and
finally over $\ell_1$, the final net result is:

$$
Tr (R^{ee} R^{ee} (\mu^{N-n} \otimes \openone)) =
\frac{1}{(1-t^{-2 }) (1-\mu)}
\eqno{(A8)}
$$

\noindent
so that

$$
Z_{\widehat{\sigma^2_1}} = \frac{1}{t-t^{-1}}
\eqno{(A9)}
$$

\noindent
which is precisely the inverse of the Alexander--Conway
polynomial of the link.

Similarly we obtain for the reverse link

$$
Z_{\widehat{\sigma^{-2}_1}} =    \frac{-1}{t-t^{-1}}
\eqno{(A10)}
$$

For more complex links or knots the calculation of the invariant
requires the knowledge of complicated combinatorial formulae.
There is however a way to overcome this difficulty, which
consist in making an integral representation of the $\delta$
``functions'' in $R$ and $R^{-1}$; namely:

$$
\delta_{r_1+r_2, r'_1 + r'_2} = \oint \frac{dz}{2 \pi i} z^{-1 +
r_1 + r_2 - r'_1-r'_2}
\eqno{(A11)}
$$

\noindent
where the contour integral is performed around the origin. Thus
we associate to each braiding matrix $R$ or $R^{-1}$ a complex
variable $z$; having done this we can perform all the sums over
the entries so that $tr (\pi (\alpha) \mu^N \otimes \cdots \otimes
\openone))$ gets converted into a multicontour integral.

Let us exemplify this method with the computation of the
invariant of the trefoil.

\[
Z (\sigma^3_1 \in B_2) = (1-\mu) \; t^{-2} \; Tr R^3 (e,e) (\mu^N
\otimes \openone)
\]

\[
Tr_{\mu}(R^3 (e,e))= \sum_{r,\ell,m} \; \; R^{m_1 m_2}_{r_1
r_2} \; R^{r_1 r_2}_{\ell_1 \ell_2} \; \; R^{\ell_1 \ell_2}_{m_1
m_2} \; \mu^{m_1}
\]

\[
= \sum_{r,l,m} \oint \frac{dz_1}{2\pi i} \frac{dz_2}{2 \pi i} \;
z_1^{-1 +r_1 + r_2 - m_1 - m_2} \; z_2^{-1 + \ell_1 + \ell_2 -
r_1 - r_2} \; \mu^{m_1}
\]

$$
\left( \begin{array}{c} r_1 \\ m_2 \end{array} \right) \; \left(
\begin{array}{c} \ell_1 \\ r_2 \end{array} \right) \; \left(
\begin{array}{c} m_1 \\ \ell_2 \end{array} \right) \;
(t-t^{-1})^{r_1 + \ell_1 + m_1 - r_2 - \ell_2 - m_2} \; \;
t^{-(r_1+\ell_1 + m_1 + r_2 + \ell_2 + m_2)}
\eqno{(A12)}
$$

Notice that we have not introduced an integral representation
for the delta $\delta_{\ell_1 + \ell_2, m_1 + m_2}$, since this
is implied by the first two deltas in (A12). The sum over $r_2,
\ell_2$ and $m_2$ can be done straigforwardly yielding:

\[
\oint \frac{dz_1 dz_2}{(2 \pi i)^2} \; z^{-1}_1 z^{-1}_2
\]

\[
\sum_{r_1} \; \left( \frac{1 + t(t-t^{-1})z_1}{t^2 z_2}
\right)^{r_1} \; \sum_{\ell_1} \left( \frac{z_1 + z_2 t
(t-t^{-1})}{t^2} \right)^{\ell_1} \; \; \sum_{m_1} \left(
\frac{\mu [z_2 + (t - t^{-1})]}{t^2 z_1} \right)^{m_1}
\]

$$
= \oint \frac{dz_1 dz_2}{(2 \pi i)^2} \frac{t^6}{(t^2 z_2 -
t(t-t^{-1})z_1-1)(t^2-z_1-z_2t (t-t^{-1}))(t^2 z_1-\mu z_2-\mu
t (t-t^{-1}))}
\eqno{(A13)}
$$

The convergence of the geometric sums in $r_1, \ell_1$ and $m_1$
impose the following restrictions on $z_1, z_2$ and $\mu$:

\[
|1 + t(t-t^{-1}) z_1| < |t^2 z_2|
\]

$$
|z_1 + z_2 t (t-t^{-1})| < |t^2|
\eqno{(A14)}
$$

\[
| \mu (z_2 + t (t-t^{-1})| < |t^2 z_1|
\]

If $\mu = 0$ we see from (A13) that the integrand has a pole at
$z_1 = 0$, integrating around this pole we get

$$
\oint \frac{dz_2}{2 \pi i} \; \; \frac{t}{(t^{-1}-t)} \; \;
\frac{1}{(z_2 - \frac{1}{t^2}) (z_2 - \frac{t}{t-t^{-1}})}
\eqno{(A15)}
$$

\noindent
and from (A14) we see that

$$
\frac{1}{|t^2|} < |z_2| < \left| \frac{t}{t-t^{-1}} \right|
\eqno{(A16)}
$$

So that finally

$$
Tr R^3 = \frac{t^2}{t^2 + t^{-2} - 1}
\eqno{(A17)}
$$

\noindent
which yields:

$$
Z ({\rm Trefoil}) \; = \frac{1}{1+ (t-t^{-1})^2}
\eqno{(A18)}
$$

\noindent
according to the conjecture (46).

We have also checked $\widehat{\sigma_1^4}$ in $B_2$ and
$\widehat{\sigma^2_1 \sigma^2_2}$ and $\widehat{(\sigma_1
\sigma^{-1}_2)^3}$ in $B_3$, obtaining the result (44).

\newpage
\section*{Appendix B}

\subsection*{Clebsch--Gordan coefficients of $U_q (h_4)$}

If $(e_1, n_1)$ and $(e_2, n_2)$ are two irreps of $U_q (h_4)$
with $e_1, e_2$ and $e_1 + e_2$ different from zero then the
tensor product $(e_1,n_1) \otimes (e_2, n_2)$ decomposes into
irreps as follows:

$$
(e_1, n_1) \otimes (e_2, n_2) = \bigoplus_{n \geq 0} (e_1 + e_2,
n_1 + n_2 + n)
\eqno{(B1)}
$$

The normalized heighest weight vector of the irrep $(e_1 + e_2,
n_1 + n_2 + n)$ is given by:

\[
|0; \; e_1 + e_2, \; n_1 + n_2 + n> = \frac{1}{[e_1 + e_2]^{n/2}}
\]

\[
\times \sum_{r \geq 0} (-1)^r [e_1]^{\frac{n-r}{2}} [e_2]^{\frac{r}{2}}
q^{- \frac{re_1}{2}} q^{\frac{n-r}{2}e_2} \left(
\begin{array}{c} n \\ r \end{array} \right)^{1/2}
\]

$$
\times |r ; \; e_1, n_1 > \otimes |n - r; \; e_2, n_2>
\eqno{(B2)}
$$

The whole CG--coefficients for the tensor product decomposition
(B1) can be found from (B2) acting with the creation operators $a^+$.

An interesting application of (B2) is to find the braiding
factors $\phi$ associated with the $R$--matrix (33) which we
define as:

$$
R^{(e_1 n_1), (e_2 n_2)}  K^{(e_1 n_1), (e_2 n_2)}_{(e_1 + e_2,
n_1 + n_2 + n)}
\eqno{(B3)}
$$

\[
= \phi^{(e_1 n_1), (e_2 n_2)}_{(e_1 + e_2, n_1 + n_2 + n)} \; \;
K^{(e_2 n_2) (e_1 n_1)}_{(e_1 + e_2, n_1 + n_2 + n)}
\]

\noindent
where $K^{12}_3$ is the CG operator: $V_3 \rightarrow V_1
\otimes V_2$, which maps the irrep ``3'' into the tensor product
``$1 \otimes 2$''.

Using (33) and (B2) one finds

$$
\phi^{(e_1 n_1), (e_2 n_2)}_{(e_1 + e_2, n_1 + n_2 + n)} \; \; =
(-1)^n q^{-(e_1 + e_2)n}
\eqno{(B4)}
$$

Putting back the factor $q^{-e_1 n_2 - e_2 n_1}$, that we
discard in the definition of $R^{e_1 e_2}$ (eq. 32), we see that
the whole braiding factor induced by the $R$ matrix is given by:

$$
(-1)^n \; \; q^{e_1 n_1 + e_2 n_2 - (e_1+e_2) (n_1+n_2+n_)}
\eqno{(B5)}
$$

We recognize in this expression the classical casimir $C_1$
in the exponential $q^{en}$, which is the analog of the braiding
factor $q^{j(j+1)}$ of the quantum group $U_q$ (SU(4)).
Similarly $(-1)^n$ is a parity factor analogous to $(-1)^j$ in SU(2).

\newpage

\section*{Appendix C}

\subsection*{$U_q (h_4)$ and Ribbon Hopf Algebras}

Any quasitriangular $A$ has an invertible element, usually
called $u$, with the property that

$$
\gamma^2 (a) = u \; a \; u^{-1} \; \; \forall  a \in A
\eqno{(C1)}
$$

The element $u$ and its inverse $u^{-1}$ can be obtained from
the universal ${\cal R}$ matrix as follows:

$$
\begin{array}{lcl}
u & = & m [\gamma \otimes id (\sigma({\cal R}))] \\
& & \\
u^{-1} & = & m [id \otimes \gamma^2 (\sigma ({\cal R}))] \end{array}
\eqno{(C2)}
$$

In the particular case of the universal $R$--matrix (30) of the
quantum group $U_q (h_4)$ we obtain:

$$
\begin{array}{lcl}
u & = & \sum_{\ell \geq 0} \frac{(q^{-1} - q)^\ell}{\ell !} \;
q^{-\ell E} (a^+)^\ell a^\ell q^{2EN} \\
& & \\
u^{-1} & = & \sum_{\ell \geq 0} \frac{(q - q^{-1})^\ell}{\ell !} \;
q^{\ell E} (a^+)^\ell a^\ell q^{-2EN} \end{array}
\eqno{(C3)}
$$

The element $u$ of the quantum deformation $U_q ({\cal G})$ of a
complex simple Lie algebra ${\cal G}$ does not in general commute with
the others elements of the algebra, however for $U_q (h_4)$ we
have from eqs (27) that $\gamma^2 = id$ so that $u$ is central.
Indeed using (C3) and (31) we obtain:

$$
u | r; \; (e, n) > = q^{2en} |r;( e,n)>
\eqno{(C4)}
$$

Similarly $\gamma (u)$ is also central and one can prove that

$$
\gamma (u) = q^{-2E} \;u
\eqno{(C5)}
$$

We are now in conditions to study wether $U_q (h_4)$ is a Ribbon
Hopf algebra. Following Resbetikhin and Turaev [2] we define a
Ribbon Hopf Algebra $(A, R, v)$ as a quasitriangular Hopf
algebra with a universal $R$ matrix and the choice of an element
$v$ such that:

$$
\begin{array}{lcl}
v & is & central \\
v^2 & = & u \gamma (u) \\
\gamma (v) & = & v \\
\varepsilon (v) & = & 1 \\
\Delta (v u^{-1}) & = & v u^{-1} \otimes v u^{-1} \end{array}
\eqno{(C6)}
$$

In the case of $U_q (h_4)$ we can easily see using eq. (C5) that
the element $v$ is given by:

$$
v = q^{-E} \; u
\eqno{(C7)}
$$

\noindent
which in the irrep $(e,n)$ takes the value

$$
v | r ; (e,n) > = q^{(2n-1)e} |r ; (e,n)>
\eqno{(C8)}
$$

The fact that $v^{-1} u = q^E$ explains the presence of the term
$q^{em}$ in the invariant (37) or the corresponding term
in the multicolored
version (49), and it is the $U_q (h_4)$ analogue of the operator
$q^{H/2}$ of the quantum group $U_q$ (SU(2)).

The value of $v$ in a given representation of a quantum group
$A$ contains some interesting information of the corresponding
conformal field theory associated to $A$. In the cases where $A
= U_q ({\cal G})$ with ${\cal G}$ a simple Lie algebra it was shown in
reference [14] that the conformal weight $\Delta_\alpha$
of a primary field of the WZW model $\hat{{\cal G}}_k$ is related to the
value $v_\alpha$ by:

$$
v_\alpha = e^{2 \pi i \Delta_\alpha}
\eqno{(C9)}
$$

\noindent
where $v_\alpha$ is the value of $v$ on the irrep $\alpha$ of
$U_q ({\cal G})$ associated to the primary field $\alpha$. If eq. (C9)
would holds true for $h_4$ it would imply that

$$
q^{(2n-1)e} = e^{2 \pi i \Delta_{(e,n)}}
\eqno{(C10)}
$$

In ref [7], where it is consider a WZW  theory based on the
supergroup GL(1,1), eq. (C10) is violated by higher order
corrections of order $1/k^2$ with $k$ the level of the
Kac--Moody supergroup GL(1,1).

In our case we do not know yet wether there is an underlying
conformal field theory associated to the quantum group $U_q (h_4)$.

\end{document}